\documentstyle[preprint,aps]{revtex}

\begin{document}
\draft
\title{Transport of one-dimensional interacting Fermions through a barrier}
\author{Q. P. Li\cite{byline}}
\address{Department of Physics and Applied Superconductivity Center\\
 University of Wisconsin-Madison, 1150 University Avenue,
Madison, Wisconsin 53706 }
\author{X. C. Xie}
\address{Department of Physics,
Oklahoma State University \\
Stillwater, Oklahoma 74078-0444 }
\date{\today}
\maketitle
\begin{abstract}
We study the transport properties of one-dimensional (1D) interacting
Fermions through a barrier by numerically calculating the Kohn charge stiffness
constant and the relative Drude weight.
We find that the transport
properties of the 1D Hubbard model are quite different from those of
the 1D spinless Fermion model. For example, the presence of the
attractive interaction between electrons in the 1D Hubbard model actually
{\em suppresses} the DC conductance, while a small repulsive
 interaction {\em enhances} the DC conductance.  These
results show that the spin degree of freedom plays an important role
in the transport properties of the 1D interacting Fermion systems.
\end{abstract}
\narrowtext

Recently, there has been increasing interest in one-dimensional (1D)
electron systems.\cite{hans,wees,scott,meir,goni,li,kane,furu} On the
one hand, the progress in ultrafine
lithographic technology has made it possible to fabricate and
experimentally study quantum
wires in which only the lowest subband is occupied, approaching the
true 1D limit; on the other hand, there is a renewed interest in the
 Luttinger liquid because of the suggestion that
the high temperature superconductors might be non-Fermi-liquid-like. It is
well known that even a
weak interaction in a 1D Fermion system will drive the system to become
a Luttinger liquid which exhibits quite different behavior from the
Fermi liquid system.\cite{lutt,hald} For example, Kane and Fisher have
studied the
transport properties in a 1D spinless Luttinger liquid with a barrier
(or a weak link). They find
that an attractive interaction   makes the system a perfect conductor while
a repulsive interaction makes it an insulator.\cite{kane}

In this paper we present some results of the transport
properties of 1D interacting Fermion systems with a barrier. We
study both the interacting spinless Fermion model and the 1D Hubbard model
with a barrier  by calculating
the Kohn charge stiffness constant using the Lanczos algorithm. In particular,
we investigate
the effect of spin on the transport properties motivated by an interesting
statement of Anderson who argued that the theory of the transport in
1D spinless Fermion may be irrelevant for a real electron system which has
both charge and spin degrees of freedom.\cite{ande} We find that our result
for the
spinless Fermion model agrees with that of Kane and Fisher qualitatively.
But our result of the 1D Hubbard model with a barrier is exact opposite
of that of the spinless Fermion model. We find that in the case of the
Hubbard model, the presence of the attractive interactions between
electrons actually {\em decreases} the Kohn charge stiffness constant
D$_{c}$ as
well as the relative Drude weight, in contrast to the 1D spinless
Fermion case where the presence of attractive interactions {\em increases}
D$_{c}$ and the relative Drude weight as long as the attractive
interaction is not large enough to cause a phase transition to a
phase-separated insulating state.
In the case of
repulsive interaction, we find that a small repulsive interaction
in the 1D Hubbard model with a weak link actually {\em increases} D$_{c}$
and the relative
Drude weight, again in contrast to the 1D spinless Fermion model where
the repulsive interaction {\em decreases} D$_{c}$.

We first study an interacting spinless Fermion model, so called
Heisenberg-Ising model, on a 1D ring with a weak link
\begin{eqnarray}
H = -\sum_{i} t_{i}(c_{i}^{\dag}c_{i+1} + h.c.)
+ V\sum_{i} n_{i}n_{i+1}.
\end{eqnarray}
Here $t_{i}=t$ for $i \neq 1$ and $t_{1}=t^{\prime}$ representing the
barrier (or weak link) site, and
V is the nearest neighbor interaction. We assume the periodic
boundary condition and thread the 1D ring with a flux $\Phi$ which is
represented by a vector potential ${\bf A} = (\Phi/L) \hat{e}_{x}$, where
L is the number of sites. We calculate numerically
the ground state energy of the
system $E(\Phi)$ as a function of flux, and consequently the
Kohn charge stiffness constant
\begin{eqnarray}
D_{c} = \frac{L}{2} \frac{d^{2}E(\Phi)}{d\Phi^{2}}|_{\Phi=0}.
\end{eqnarray}
$D_c$ is a measure of the electronic conductivity of
the system, and it vanishes if the system is an
insulator. Kohn first suggested that the charge stiffness
constant can be used as a quantitative measure of the Mott metal-insulator
transition.\cite{kohn,mill,poil,li2}

Physically, the charge stiffness constant $D_{c}$ is just the
Drude weight in the real part of the optical conductivity
$\sigma_{1}(\omega)$ in the long wavelength limit,
\begin{eqnarray}
\sigma_{1}(\omega) = 2\pi e^{2} D_{c} \delta(\omega)
+ \sigma_{1}^{reg}(\omega),
\end{eqnarray}
here we have set $\hbar=1$. Using second order
perturbation theory, one finds,\cite{shas}
\begin{eqnarray}
D_{c} = \frac{1}{2L}<-T> - \frac{1}{L}\sum_{m\neq 0}
\frac{|<0|j_{x}|m>|^{2}}{E_{m}-E_{0}},
\end{eqnarray}
where $<-T>$ is the expectation value of the kinetic energy in the
ground state. $\sigma_{1}(\omega)$ satisfies the well-known f-sum
rule,
\begin{eqnarray}
\int_{-\infty}^{\infty} \sigma_{1}(\omega) d\omega =
\frac{\pi e^{2}}{L} <-T>.
\end{eqnarray}

We use the Lanczos algorithm to calculate the ground state energy. For a
14-site ring with 7 particles, the dimension of the Hilbert space is about
3400 and the Hamiltonian matrix has about 30,000 non-zero elements which can
be handled easily on a workstation. We first consider a 1D ring with the
lattice translational symmetry,
i.e., $t^{\prime}=t$, in which most of the properties are known.
This well-studied case provides us with a test example which
we can check our computer program against and get some feelings about the
finite size effect, etc.  In Fig. 1a, we show the calculated ground
state energy $E(\Phi)$ as a function
of flux $\Phi$ for a 14-site ring with 7 and 6 particles, respectively.
Obviously, $E(\Phi)$ is a periodic function of $\Phi$ with a period
$\Phi_{0}=hc/e$. For the 7 particle system, the energy minimum is located at
$\Phi=0$ as expected. This is true as long as the number of particles is an
odd integer. But when there are even number of particles on a ring, such as in
the 6-particle case, the minimum of $E(\Phi)$ is located at $\Phi=\Phi_{0}/2$,
signaling that the system prefers a spontaneous persistent current which
would generate a flux of $\Phi_{0}/2$. This is the well-known even-odd
finite size effect.\cite{douc} Fig. 1b shows the calculated Kohn
charge stiffness constant $D_{c}$ as a function of the nearest neighbor
interaction V with L=14, N=7, and $t^{\prime}=t$. One can see that on the
repulsive interaction side, $D_{c}$
approaches zero quickly as the repulsive interaction increases, signaling
a transition to the Mott-insulator in this half-filling case. On the
attractive interaction side, $D_{c}$ also start to decrease, even faster
than the repulsive case, at around
$V=-2t$. This is the transition to a phase-separated insulator state.
On the repulsive interaction side, it is
well known that $D_{c}$ and the effective mass have a discontinuous jump at
 $V=2t$ in the thermodynamic limit.\cite{shas} As $V \rightarrow 2t-0$,
D$_{c}$ approaches a nonzero
value 1/4, while $D_{c}=0$ as $V \rightarrow 2t+0$.
Our finite size calculation yields
$D_{c} \simeq 0.285$ at $V=2t$. At V=0, D$_{c}$ in the thermodynamic
limit is $1/\pi \simeq 0.3183$. Our calculation gives $D_{c} \simeq 0.3210$.
The finite size calculation usually overestimates D$_{c}$ because of the
periodic boundary condition.

In Fig. 2a, we show $D_{c}$ as a function of V for a 5-particle 16-site
system with $t^{\prime}=$t, 0.75t, and 0.25t, respectively. Fig. 2b shows the
relative Drude weight (RDW) which is defined as the ratio of
the Drude weight $D_{c}$ and the total spectral weight $A_{tot} =
\int_{-\infty}^{\infty} \sigma_{1}(\omega) d\omega /2\pi e^{2}$
 for the same system. The first thing we can see from
Fig. 2 is that the presence of a weak link does not seem to affect the
phase-separation transition much.
The weak link actually  makes the phase-separation-transition sharper in
a finite size system.   The
transition is depicted better in the relative Drude weight plot (cf. Fig. 2b).
It is obvious that any theory which assumes a uniform phase such as that of
Kane and
Fisher\cite{kane} is only applicable above the threshold $V_{c}=-2t$. Below
$V_{c}$, $D_{c}$ is zero  and the system is an insulator in the thermodynamic
limit. The second thing we observe from Fig. 2 is that the attractive
interaction does enhance the conductivity, especially when $t^{\prime}/t$
is small, that is quite different from the result of 1D Hubbard model which
we'll
discuss later, while the repulsive interaction suppress the conductivity. This
is in qualitative agreement with the analytical studies of Kane and
Fisher\cite{kane}, and Apel and Rice.\cite{apel}
However, due to the finite size effect and the phase-separation transition at
$V_{c}=-2t$,  presently we are not able to confirm
Kane and Fisher's specific prediction\cite{kane} that any repulsive
interaction will drive the system to insulating while any attractive
interaction will make the system a perfect conductor. We find
that $D_{c}$ and the
relative Drude weight remains finite even for very large positive V in a
finite system.
The third point we learn from Fig. 2 is that when $t^{\prime}=t$, the
Drude weight almost exhausts all the spectral weight $A_{tot}$ as long
as V is above the phase-separation threshold $V_{c}=-2t$. The relative
Drude weight drops quite rapidly once V is decreased below $V_{c}$.
When $t^{\prime}<t$, the RDW is usually smaller than the $t^{\prime}=t$
case (cf. Fig. 2b). This is due to the pinning of the charge density
wave by the weak link, which not only reduces $D_{c}$, but also reduces
the RDW.

Next we consider the Hubbard model on a 1D ring with a weak link,
\begin{eqnarray}
H = -\sum_{i\sigma} t_{i}(c_{i\sigma}^{\dag}c_{i+1\sigma} + h.c.)
+ U\sum_{i} n_{i\uparrow}n_{i\downarrow}.
\end{eqnarray}
Here again $t_{1}=t^{\prime}$ and $t_{i}=t$ for $i\neq 1$, and $\sigma =
\uparrow, \downarrow$ is the spin index. In Fig. 3 we
show the calculated D$_{c}$ and the relative Drude weight as functions
of the interaction U for $t^{\prime}=$t, 0.5t, and 0.1t, respectively.
We find that the transport properties of the 1D hubbard model are quite
different from those of the spinless Fermion model. In Hubbard model, the
presence of the attractive interaction between electrons actually
{\em decreases} D$_{c}$ and RDW (cf. Fig. 3), in contrast to the spinless
Fermion model (cf. Fig. 2 and Refs.\cite{kane,apel}). The result of the
repulsive interaction is equally surprising.
The presence of a small repulsive interaction ($U > 0$) increases D$_{c}$
and RDW. So the maximum of D$_{c}$ actually shifts from U=0 for
$t^{\prime}=t$ to some small positive $U_{0}(t^{\prime})$ for
$t^{\prime}<t$ (cf. Fig. 3). $U_{0}(t^{\prime})$ increases as $t^{\prime}$
decreases. For $t^{\prime} = 0.1t$, $U_{0} \simeq t$. Since the only major
difference between the Hubbard
model and the spinless Fermion model is the spin, it is reasonable to
suspect that the spin does play a very important role in the transport
of 1D interacting Fermions, which is in agreement with Anderson's
conclusion.\cite{ande}

The fact that in Hubbard model, D$_{c}$ decreases as
one increases the strength of the {\em attractive} interaction can
be understood from  the
following physical picture. In the
attractive interaction region, electrons form pairs which can be
treated as hard-core bosons,\cite{note}, and these bosons can hop via virtual
ionization.
A straight forward second-order perturbation calculation yields that the
pair hopping amplitude is $2t^{2}/|U|$ and the nearest neighbor
repulsion between the bosons
is also $2t^{2}/|U|$ in the $t^{\prime}=t$ case. Physically, this
is easy to understand because the virtual ionization to a nearby site
lowers the energy, but it is only possible if the nearby site is
empty.\cite{nozi}

Comparing our results with that of Ref. 7 and 8, we find that although
our results in the spinless case is in qualitative agreement with their
results,  there is significant discrepancy between our results and that
of  Ref. 7 and 8 in the spinful Fermion case. The Hubbard model
we studied has SU(2) symmetry, so $g_{\sigma}$ is fixed at 2 (we use the
same notation $g_{\sigma}$ as in Ref. 7). This means,
according to Ref. 7 and 8, that the spin part should behave like the
noninteracting case while the charge part should act like the spinless
Fermion. This is in disagreement with our finding that the presence of
a small repulsive interaction in Hubbard model with a week link actually
increases DC conductance which is quite different from the results of
spinless Fermions. One possible explanation of our results is that the
presence of repulsive interaction increases the antiferromagnetic
correlation which may enhances the transport across the weak link. Some of
these physics is apparently missing in Ref. 7 and 8. For example, Ref. 8
starts from a model which only has interaction $g_{2}$ and completely
neglects the $2k_{F}$ component $g_{1}$. Here $g_{1}$ and $g_{2}$ are
standard ``g-ology'' notations as in Ref. 9.

Finally we mention that although all of the results shown above are
for $t^{\prime} \leq t$, we have also studied the case of
$t^{\prime} > t$. In this case, we find that D$_{c}$ decreases as
$t^{\prime}$ is increased from t, in agreement with the picture that the
$t^{\prime} \neq t$ link (either $t^{\prime} < t$ or
$t^{\prime} > t$) serves as a pinning center of the charge density
wave in the repulsive interaction region.

In conclusion, we have studied the transport properties of both the
1D spinless Fermion model and the  Hubbard model
with  a weak link by calculating the Kohn charge stiffness
constant D$_{c}$ and the relative Drude weight using Lanczos algorithm.
We find that the presence of the
attractive interaction between electrons in 1D Hubbard model actually
{\em suppresses} the DC conductivity, which is in contrast to the 1D
spinless Fermion model where the attractive interaction {\em enhances}
the DC conductivity as long as the attractive is not large enough to
cause a phase transition to a phase-separated insulating state. In the
repulsive interaction region, we find that a small repulsive interaction
in 1D Hubbard model actually {\em increases} D$_{c}$ and the relative
Drude weight, again in contrast to the 1D spinless Fermion model where
the repulsive interaction {\em decreases} D$_{c}$. These
results show that the spin degree of freedom plays an important role
in the transport properties of the 1D interacting Fermion systems.

\acknowledgements

It is a pleasure to thank S. Das Sarma, Steve Girvin, Song He,
Ben Hu, Robert Joynt, G. Kotliar, Qian Niu, C. Stafford, Shuan Tang,
and Nandini Trivedi  for valuable
conversations. QPL would like to thank the hospitality of the Aspen
Center for Physics where part of the work was done. This work was supported
in part by the NSF
through Grant No. DMR 9214739, and by the Electric Power Research
Institute. We also acknowledge support from San Diego Supercomputer
Center.

\begin{figure}
\caption{(a). Shows the calculated ground state energy $E(\Phi)$ of the
Heisenberg-Ising model as a function
of the  flux $\Phi$ for N=7 (solid) and 6 (dashed), respectively. Here L=14,
V=2t, and $t^{\prime}=t$. The unit of $E$ is t. (b). Shows the calculated
$D_{c}$ as a function of interaction V for the N=7 system. }
\label{half}
\end{figure}

\begin{figure}
\caption{Shows the calculated (a)
$D_{c}$ and (b) the relative Drude weight as a function of interaction V of a
5-spinless-Fermion system on a  16-site 1D ring  with
$t^{\prime}$=t (solid), 0.75t (dashed), and 0.25t (dot-dashed), respectively.}
\label{dc}
\end{figure}

\begin{figure}
\caption{Shows the calculated (a)
$D_{c}$ and (b) the relative Drude weight of a 1D Hubbard model as a function
of interaction U of a
6-Fermion system on a  8-site  ring  with
$t^{\prime}$=t (solid), 0.5t (dashed), and 0.1t (dot), respectively. The
total spin $S_{z} = 0$.   }
\label{hdc}
\end{figure}


\begin{references}
\bibitem{note} In 1D Hubbard model with an attractive interaction, a pair
of electrons always form a bound state as long as the attractive interaction
is nonzero. See Ref.\cite{li2}.
\end{references}
\end{document}